\documentclass[%
reprint,
amsmath,amssymb,
aps,
pre,
twocolumn,
]{revtex4-2}

\usepackage{tikz}
\usepackage{graphicx}% Include figure files
\usepackage{dcolumn}% Align table columns on decimal point
\definecolor{red}{rgb}{0.968627450980392,0.137254901960784,0.0470588235294118}
\definecolor{blue}{rgb}{0.00784313725490196,0.192156862745098,0.709803921568627}

\begin{document}

\preprint{APS/123-QED}

\title{Statistics of a 2D immersed granular gas magnetically forced in volume}% Force line breaks with \\

\author{Jean-Baptiste Gorce}
\email{Jean-Baptiste.Gorce@u-paris.fr}
% \altaffiliation[Also at ]{Physics Department, XYZ University.}%Lines break automatically or can be forced with \\
\author{Eric Falcon}%
\affiliation{Universit\'e Paris Cit\'e, CNRS, MSC, UMR 7057, F-75013 Paris, France}%

%\collaboration{MUSO Collaboration}%\noaffiliation
%
%\author{Charlie Author}
% \homepage{http://www.Second.institution.edu/~Charlie.Author}
%\affiliation{
% Second institution and/or address\\
% This line break forced% with \\
%}%
%\affiliation{
% Third institution, the second for Charlie Author
%}%
%\author{Delta Author}
%\affiliation{%
% Authors' institution and/or address\\
% This line break forced with \textbackslash\textbackslash
%}%

\date{\today}% It is always \today, today,
% but any date may be explicitly specified

\begin{abstract}

We present an experimental study of the dynamics of a set of magnets within a fluid in which a remote torque applied by a vertical oscillating magnetic field transfers angular momentum to individual magnets. This system differs from previous experimental studies of granular gas where the energy is injected by vibrating the boundaries. Here, we do not observe any cluster formation, orientational correlation and equipartition of the energy. The \textcolor{black}{magnets}' linear velocity distributions are stretched exponentials, \textcolor{black}{similar to} 3D boundary-forced dry granular gas systems, but the exponent does not depend on the number of magnets. The value of the exponent of the stretched exponential distributions is close to the value of 3/2 previously derived theoretically. Our results also show that the conversion rate of angular momentum into linear momentum during the collisions controls the dynamics of this homogenously-forced granular gas. \textcolor{black}{We report the differences between this homogeneously-forced granular gas, ideal gas, and nonequilibrium boundary-forced dissipative granular gas.}

\end{abstract}

\maketitle

\paragraph*{Introduction.\textemdash}Granular gases are many-particle systems in which individual particles undergo random motions, and whose dynamic \textcolor{black}{differs} from molecular gases due to the energy loss during collisions \cite{poschel2001,poschel2003}. Granular gases are complex out-of-equilibrium systems and show unique properties compared to molecular gases as they can violate the time-reversal symmetry \cite{poschel2002,shaw2007} and show non-equipartition of energy \textcolor{black}{ \cite{goldshtein1995,feitosa2002,galvin2005,brilliantov2007,gayen2011}}. The inelastic collisions between the particles imply that granular gases are dissipative and require a constant input of energy to compensate for the loss of kinetic energy. They are usually boundary-driven, for example by vibrating the boundary of the container.

The theoretical framework of granular gases assumes a homogeneous forcing and a high energy tail velocity distribution of the particles $P\left(v\right) \sim \exp\left(-av^{3/2}\right)$, where $v$ is the dimensionless velocity and $a$ a constant involving the restitution coefficient $\epsilon$ \cite{van1998}. \textcolor{black}{Many} experimental measurements have shown that the value of the exponent in the stretched exponential depends on the number of particles \cite{esipov1997,losert1999,olafsen1999,rouyer2000,kudrolli2000,luding2003,van2004,wang2009,tatsumi2009,scholz2017}. The dissipation of the kinetic energy in these boundary-driven granular gases leads to spatial inhomogeneity such as clustering of monodispersed particles \cite{goldhirsch1993,kudrolli1997,falcon1999,falcon1999b,opsomer2011,mitrano2012,noirhomme2018}, and segregation of bidisperse particles \cite{hsiau1996,jenkins2002,schroter2006,garzo2006,garzo2011,opsomer2017}.

Recently, a new experimental forcing technique was developed to inject energy directly into the volume of a granular gas instead of at the boundaries \cite{falcon2013}. A pair of Helmholtz coils generates a vertical oscillating magnetic field which transfers kinetic energy to magnetic stirrers by imposing a magnetic torque. In the gas-like regime, $N$ magnets initially sitting at the bottom of a \textcolor{black}{container} receive angular momentum, which is converted into linear momentum when the magnets collide with the side, top boundaries, or other magnets. No clustering is observed in this granular gas. The equation of state is measured in three dimensions (3D) using an accelerometer clamp on the top lid \textcolor{black}{and} differs from the equation of state of molecular gases by a geometric correction \cite{falcon2013}. The velocity statistics have an exponential tail independent of the number of the magnets \cite{falcon2013}. When the particles are immersed in a liquid medium, the random motions of the magnets within the liquid reservoir generate hydrodynamic turbulence \cite{cazaubiel2021,gorce2022}.

However, the relationship between the dynamics of the magnets and the way the energy is injected is unclear because few experimental studies focused on measuring the angular velocities of granular gases. In boundary-driven granular gases, the angular velocity distributions are either stretched exponential \cite{nichol2012,harth2013}, Gaussian \cite{schmick2008,scholz2017} or non-Gaussian in 2D \cite{jantzi2020}. Here, we investigate the dynamics of a gas of magnets using Lagrangian tracking techniques to understand how the magnets convert their angular momentum into linear momentum and how the kinetic energy is distributed between the degree of freedom in the gas. 

\begin{figure*}[t!]
\centering
\begin{tikzpicture}

\node at (-5,-0.3) { \includegraphics[width=15.5cm]{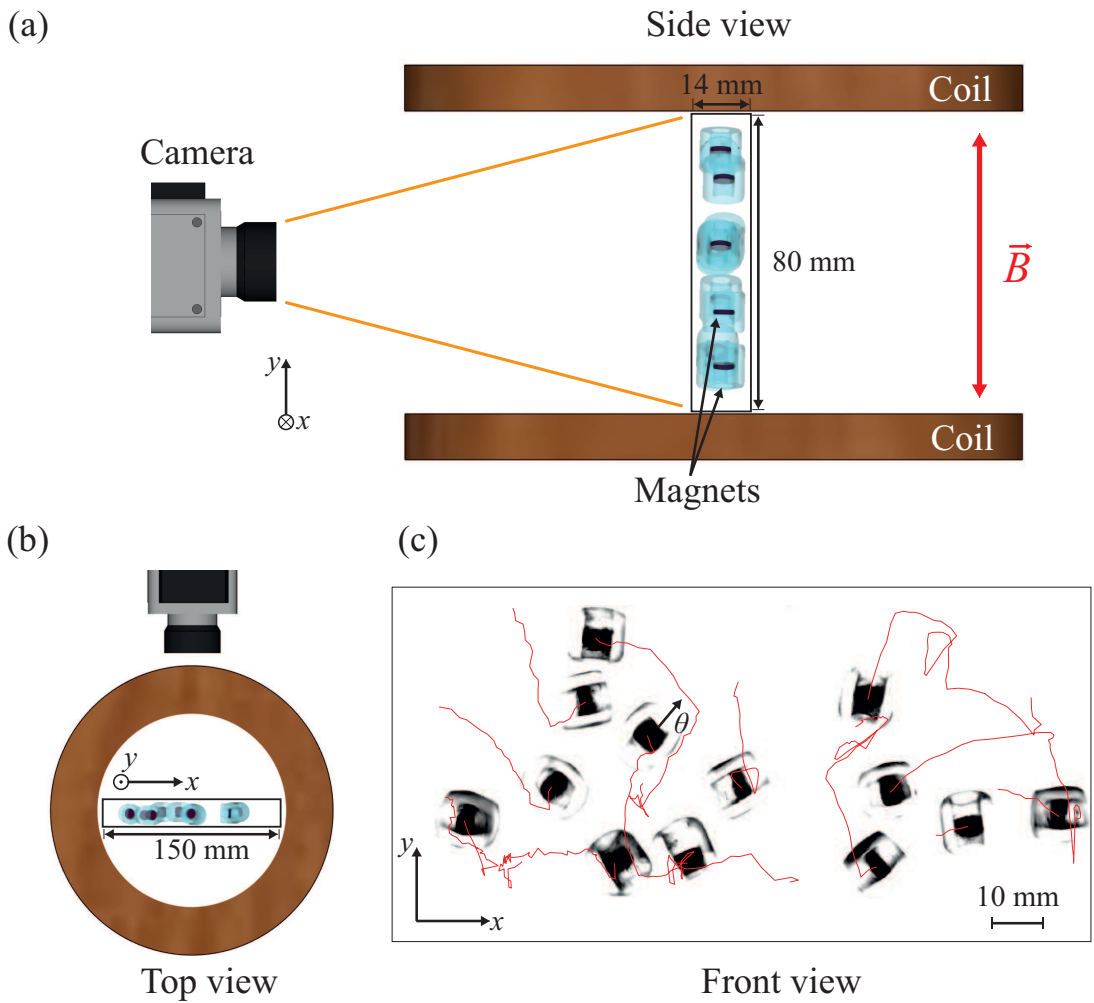}};

\end{tikzpicture}
\caption{(a,b) Schematic of the experimental setup for the 2D granular gas of magnets immersed in water and remotely energized by a vertical oscillating magnetic. The vertical magnetic field is generated by a pair of Helmholtz coils. The magnets are illuminated from the rear by a LED panel, and a high-speed camera records time series of images. (c) Trajectories of $N=13$ magnets \textcolor{black}{($\phi=0.06$)} in the fluid reservoir (red). The black cores are neodymium magnets surrounded by transparent Plexiglas shells.}
\label{sketch}
\end{figure*}

\paragraph*{Experimental setup.\textemdash}Figures \ref{sketch}a and \ref{sketch}b show a schematic of the experimental setup. The fluid container is a quasi-2D transparent Plexiglas container of dimensions $15 \times 1.4 \times 8$ cm\textsuperscript{3} and a volume $V=168$ cm\textsuperscript{3} filled with distilled water. The aim of studying the motions of the magnets in a quasi-2D liquid reservoir is to focus solely on 3 degrees of freedom which are the horizontal and vertical coordinates $(x,y)$ and the angle $\theta$ between the horizontal axis and the magnet dipolar moment. The fluid container is fitted within a pair of Helmholtz coils with an inner diameter of 18 cm and an outer diameter of 40 cm. A sinusoidal current is supplied to the coils pair via a power amplifier (Qualitysource PA 2000AB), and a waveform generator (Agilent 33220A) controls the intensity $B \in \left[0.0135,0.0216\right]$ T and frequency $f_B \in \left[5,50\right]$ Hz of the applied \textcolor{black}{vertical} magnetic field. The magnets are neodymium disks encapsulated in cylindrical shells of a diameter of 1 cm, \textcolor{black}{a} height of 1 cm \textcolor{black}{, and a} of volume $V_m=0.78$ cm\textsuperscript{3}. The Plexiglass shell aims to reduce the dipolar interaction between adjacent magnets \cite{falcon2013,falcon2017}. The number of magnets is $N \in \left[2,50\right]$, which corresponds to the volume fractions $\phi=N V_m/V \in \left[0.01,0.25\right]$. The mass of each magnet is $m=1$ g, and its moment of inertia is $I=0.14$ g\,cm\textsuperscript{2}. \textcolor{black}{The magnets have a density of $\rho=1.28$ g/cm\textsuperscript{3} and are immersed in water to decrease the effect of gravity. The dynamic is, therefore, closer to 3D boundary-forced granular gases in microgravity \cite{aumaitre2018} than to dry granular gases}. A high-speed camera, (Phantom v10 2 Mpixel - 1 kHz), records a time series of images \textcolor{black}{and Fig. \ref{sketch}c shows the erratic trajectories of the magnets in the liquid container} (see also Supplemental Material \cite{suppmat}).

\begin{figure*}
\centering
\begin{tikzpicture}

\node at (4.6,-.4) { \includegraphics[width=17cm]{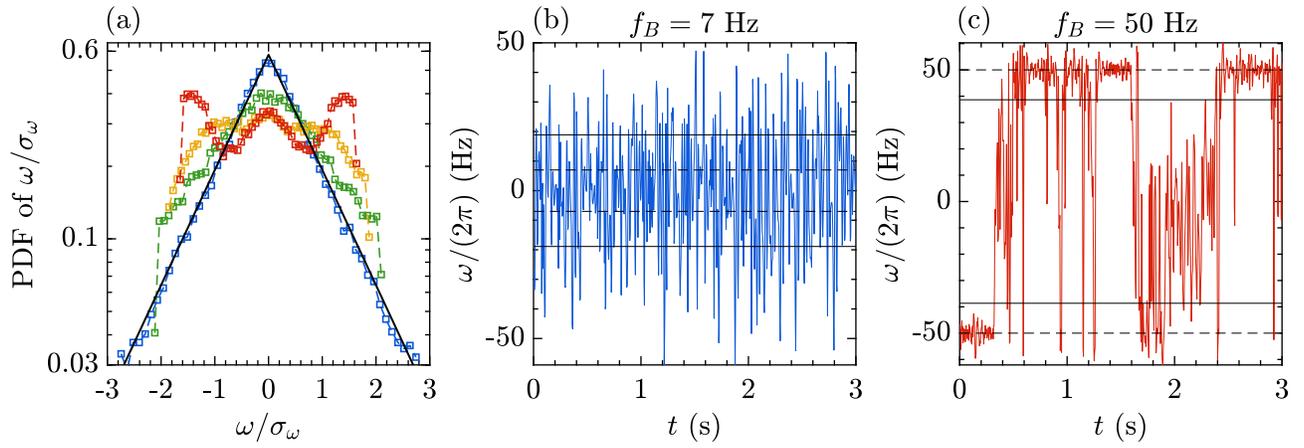}};

\end{tikzpicture}
\caption{(a) Probability density functions (PDFs) of the rescaled angular velocity $\omega/\sigma_\omega$ for different frequencies $f_B$ of the oscillating magnetic field: 7, 20, 30, and 50 Hz (cold to hot colors). The number of magnets is equal to $N=25$ \textcolor{black}{($\phi=0.12$)} and the intensity of the magnetic field to $B=0.0162$ T. The equation of the solid line is $y=a_\omega e^{-b_\omega x}$, with $a_\omega=0.58$ and $b_\omega=1.1$. (b) Temporal signal of the angular velocity $\omega/2\pi$ of a magnet for $f_B=7$ Hz. (c) Temporal signal of the angular velocity $\omega/2\pi$ of a magnet for $f_B=50$ Hz. In both (b,c), the solid lines represent the standard deviation $\sigma_\omega$ of the signal, and the dashed lines represent the frequency of the magnetic field $f_B$. }
\label{rotk}
\end{figure*}

\paragraph*{Angular velocity distributions.\textemdash} Assuming the magnets are in a liquid medium and at a high Reynolds number, the injection of energy to the magnets can be mathematically expressed using the equation of the angular momentum,

\begin{equation}
I\ddot \theta = -\lambda \dot \theta | \dot \theta |+\text{m} B \sin \left( \omega_B t\right) \sin \theta \hspace{0.3cm}, \hspace{0.3cm} \dot \theta= d\theta/dt
\label{eqmagnets}
\end{equation}

\noindent where $I$ is the moment of inertia of the magnet, $\theta$ is the angle between the vertical magnetic field and the magnetic moment \text{m} of a magnet, $\lambda=\pi \rho C_rR^5/I = 0.007$ is the damping coefficient, $ B$ the intensity and $\omega_B=2\pi f_B$ the angular frequency of the magnetic field. $C_r = 0.01$ is the coefficient of rotational drag, its value is assumed \textcolor{black}{to be equal to the one of a} sphere of radius $R$ \cite{falcon2017}.

We first measure the distributions of the angular velocity of the magnets as a function of the frequency of the magnetic field $f_B$. We define $\omega = \dot \theta$ as the angular velocity of a magnet and its standard deviation as $\sigma_\omega=\sqrt{\left<\omega^2\right>_{t,N}}$. Figure \ref{rotk}a shows the rescaled angular velocity distributions of $\omega/\sigma_\omega$ for different frequencies $f_B$. In this graphic, one can observe that the shape of \textcolor{black}{the} distributions strongly depends on $f_B$, similar to the one of a vibration-driven disk \cite{guan2021}. For a low $f_B$, the angular velocity is erratic and its probability density function is well fitted by a stretched exponential. 

\textcolor{black}{The angular velocity distribution is different from the stretched exponential distributions measured in boundary-driven rod-shaped grains in microgravity \cite{harth2013} or the Gaussian distribution observed for rolling magnetic spheres \cite{schmick2008}.} The distribution of the angular velocity of the granular gas strongly depends on the forcing frequency. Indeed, a sharp transition is observed when the frequency of the magnetic field is increased above 20 Hz, as illustrated by the shape of yellow and red curves in Fig. \ref{rotk}a. \textcolor{black}{In particular, two humps are observed in the distribution of the angular velocity at $\omega/\sigma_\omega \approx \pm 1.5$ for $f_B=50$ Hz. Those humps have also been reported in numerical studies \cite{gayen2008,gayen2011}. The humps measured in the present manuscript are linked to the synchronization of the magnets, while the humps observed in \cite{gayen2008,gayen2011} depend on the roughness of the particles.}

The erratic rotations observed at low $f_B$ become time-dependent rotations, with reversals, at high $f_B$, which is illustrated by the temporal signals of the angular velocity in Figs. \ref{rotk}b and \ref{rotk}c. Figures \ref{rotk}b and \ref{rotk}c show the influence of the magnetic field frequency $f_B$. At low frequency $f_B$, the non-deterministic behavior of the angular velocity is observed in Fig. \ref{rotk}b. At high frequency $f_B$, Fig. \ref{rotk}c illustrates that the angular velocity is sometimes equal to the frequency of the magnetic field (dashed lines). One can understand this transition using Eq. \eqref{eqmagnets} with $\lambda=0$. If the frequency of the magnetic field $f_B$ is high, the ratio of the dipolar magnetic energy $\mathrm m B$ and the rotational kinetic energy $I \omega_B^2/2$ becomes small. For $f_B=50$ Hz and $B=0.0162$ T, this ratio is equal to $0.02$. Therefore, one can approximate $\ddot \theta \approx 0$ for $f_B=50$ Hz, leading to time-dependent rotation of the angular coordinate $\theta = C_1 t + C_2 $, where $C_1$ and $C_2$ are either positive or negative constants. When the ratio of the dipolar magnetic energy $\text m B$ and the rotational kinetic energy $I \omega^2/2$ is of order unity, the angular velocity is erratic \cite{croquette1981,meissner1986}. This ratio is equal to $0.78$ for $f_B=7$ Hz and $B=0.0162$ T, which explains why the angular velocity is erratic. Such measurements emphasize the influence of rotational inertia on the magnets' dynamics. If the rotational inertia is low compared to the magnetic torque, the magnets are more likely to follow the externally imposed magnetic, and the rotations are erratic. However, high inertia implies that the magnets persist in their rotation while the magnetic torque constantly changes sign. This explains the difference in the shape of the probability density functions observed in Fig. \ref{rotk}a.

\begin{figure*}[t!]
\centering
\begin{tikzpicture}

\node at (4.6,-.4) { \includegraphics[width=12.5cm]{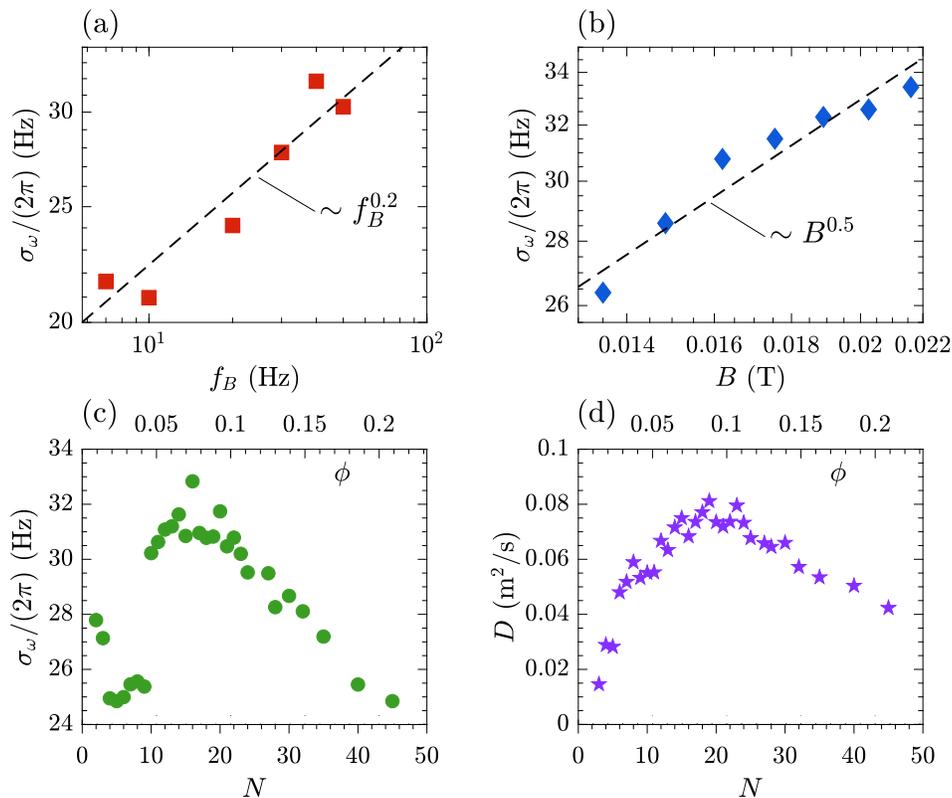}};

\end{tikzpicture}
\caption{(a) Standard deviation of the rescaled angular frequency $\sigma_\omega/\left(2\pi\right)$ as a function of the frequency of the oscillating magnetic field $f_B$ for $B=0.0162$ T and $N=25$ \textcolor{black}{($\phi=0.12$)}. The dashed line represents a power law fit of  $\sigma_\omega/\left(2\pi\right)$. (b) Standard deviation of the angular velocity  $\sigma_\omega/\left(2\pi\right)$ as a function of the intensity of the oscillating magnetic field $B$ for $f_B=50$ Hz and $N=25$ \textcolor{black}{($\phi=0.12$)}. The dashed line represents a power law fit of  $\sigma_\omega/\left(2\pi\right)$. (c) Standard deviation of the angular velocity  $\sigma_\omega/\left(2\pi\right)$ as a function of the number of magnets $N$ \textcolor{black}{or the volume fraction $\phi$}. (d) Diffusion constant $D$ of the magnets as a function of the number of magnets $N$ \textcolor{black}{or the volume fraction $\phi$}. In both (c,d) the values of the control parameters are $B=0.0162$ T and $f_B=50$ Hz.}
\label{rotk2}
\end{figure*}

\paragraph*{Rotational kinetic energy.\textemdash} Measurements of the standard deviation of the angular velocity $\sigma_\omega$ suggest that the relationship between the rotational kinetic energy and the control parameters $f_B, B$, and $N$ is complex. As shown in Figs. \ref{rotk2}a and \ref{rotk2}b, $\sigma_\omega$ is proportional to $f^{0.2} B^{0.5}$ for our range of parameters and illustrates that the energy injected into the magnets is an increasing function of the energy stored in the pair of coils.

While $\sigma_\omega$ is monotonically increasing as a function of $f_B$ and $B$, the dependence on $N$ is nontrivial. Indeed, Fig. \ref{rotk2}c illustrates that $\sigma_\omega$ increases up to a maximum value at $N=16$, which corresponds to a volume fraction $\phi=0.07$ and then decreases. We can understand the regime $N<16$ by assuming that the number of collisions between the magnets increases and, consequently, increases the fluctuations of the angular velocity. When $N$ is larger than 16, the steric effects within the reservoir become important, and the high number of collisions decreases the fluctuations of the angular velocity, thus leading to a decrease in the angular kinetic energy.

To evaluate the influence of the steric effect on the gas' dynamics, one can measure the diffusion constant $D$ defined by $\left<\left|\mathbf r_i\left(t\right)-\mathbf r_i\left(0\right) \right|^2\right>_N=4Dt$, where $\mathbf r_i(t)=x_i(t)\mathbf x+y_i(t)\mathbf y$ are the coordinates of the magnet $i$ at a time $t$. Figure \ref{rotk2}d illustrates that the diffusion constant $D$ increases when $N$ is smaller than 16, then decreases when the number of magnets $N$ is larger than 16 ($\phi=0.07$) and stresses the role of the steric effect in the gas when $N$ is larger than 16. The shape of the curves also indicates that one can link the rotational kinetic energy to the diffusion of the magnets in the reservoir. Therefore, one can assume that the collisions between the magnets control the conversion of the rotational kinetic energy into linear kinetic energy.

\begin{figure*}[t!]
\centering
\begin{tikzpicture}

\node at (4.6,-.4) { \includegraphics[width=17.5cm]{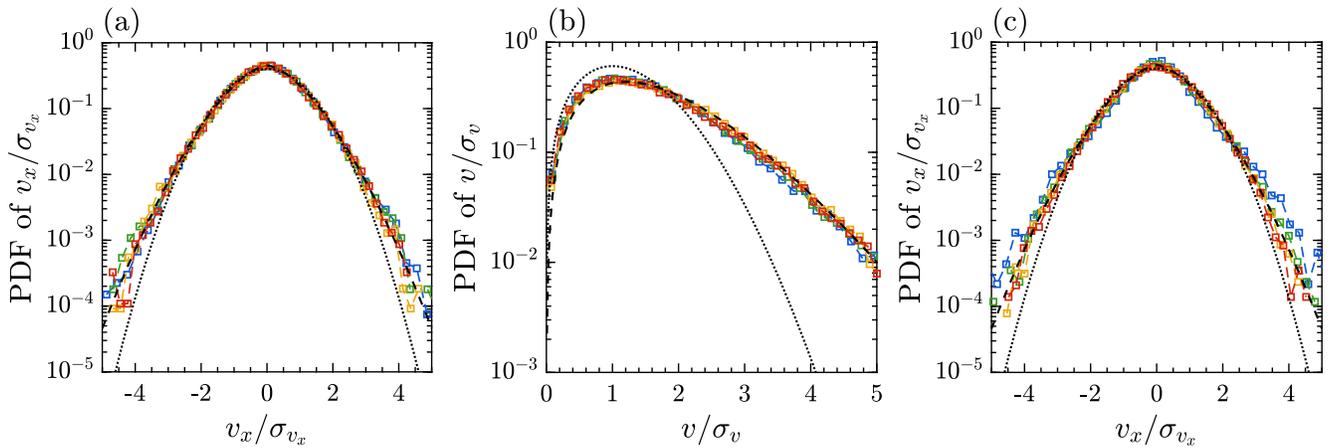}};

\end{tikzpicture}
\caption{(a) Probability density functions (PDFs) of the rescaled horizontal velocity $v_x/\sigma_{v_x}$ for different frequencies of the magnetic field $f_B$: 7, 10, 40, and 50 Hz (cold to hot colors). The equation of the dashed line is $y=a_{v_x}e^{-b_{v_x}x^{1.6}}$, with $a_{v_x}=0.45$ and $b_{v_x}=0.7$ and the dotted line represents a Gaussian distribution. (b) PDFs of the rescaled modulus of the velocity $v/\sigma_{v}$ for different frequencies of the magnetic field $f_B$. The colors are the same as in (a). The equation of the dashed line is $y=a_{v}x^{1.3}e^{-b_{v}x^{1.35}}$, with $a_{v}=0.9$ and $b_{v}=0.75$ and the dotted line represents a Rayleigh distribution. \textcolor{black}{In both (a,b), the control parameters are $B=0.0162$ T and $N=25$.} (c) PDFs of the rescaled horizontal velocity $v_x/\sigma_{v_x}$ for a different number of magnets $N$: 10, 20, 30, and 40 (cold to hot colors). The corresponding volume fractions $\phi$ are equal to 0.045, 0.09, 0.135, and 0.18. The equations of the dashed and dotted lines are the same as in (a).}
\label{link}
\end{figure*}

\paragraph*{Linear velocity distributions.\textemdash}Since the dynamics of the magnets are driven by the externally imposed magnetic field, it is, therefore, important to estimate how the angular velocity is converted into linear velocity. Figure \ref{link} illustrates that the probability density functions of the horizontal $v_x$ velocities of the magnets are very well fitted by the expression

\begin{equation}
P(v_{x})= a_{v_x} \exp\left[-b_{v_x} \left(\frac{ v_{x}}{\sigma_{v_{x}}} \right)^{1.6} \right]
\label{fit}
\end{equation}

\noindent with $\sigma_{v_x}=\sqrt{\left<v_x^2\right>_{t,N}}$, $a_{v_x}=0.45$ and $b_{v_x}=0.7$.

\textcolor{black}{The stretched exponential distributions are consistently observed for an increasing value of the magnetic field frequency $f_B$ as shown in Fig. \ref{link}a. No transition in the shape of the distributions of the linear velocity is measured, conversely to the shape of the distributions of the angular velocity.} This implies that the frequency of the magnetic field $f_B$ solely changes the standard deviation of the velocity without affecting the shape of the probability density functions. Measurements of the vertical $v_y$ velocities are also very well fitted by the same expression\textcolor{black}{, stressing the absence of the effect of gravity}. Stretched exponential distributions have been reported in different granular systems, but are different from the Gaussian distribution observed in ideal gases because of the dissipation and inelastic collisions in out-of-equilibrium systems. \textcolor{black}{As shown in Fig. \ref{link}c, the exponent 1.6 in the stretched exponential is independent of $N$ and is close to the theoretical value 3/2 \cite{van1998}. This differs from many experimental studies of 3D boundary-driven dry granular gases in which the value of this exponent depends on $N$ \cite{esipov1997,losert1999,olafsen1999,rouyer2000,kudrolli2000,luding2003,van2004,wang2009,tatsumi2009,scholz2017}.}  

Measurements of the probability density functions of the modulus of the velocity $v=\sqrt{v_x^2+v_y^2}$ in Fig. \ref{link}b illustrate that the distribution of $v$ is very well fitted by the expression

\begin{equation}
P(v)= a_v \left(\frac{v}{\sigma_{v}}\right)^{1.3} \exp\left[-b_v \left(\frac{ v}{\sigma_{v}} \right)^{1.35} \right]
\label{fit}
\end{equation}

\noindent with $\sigma_v=\sqrt{\left<v^2\right>_{t,N}}$, $a_v=0.9$ and $b_v=0.75$. Equation \eqref{fit} differs from the Rayleigh distribution observed in 2D ideal gases, which is likely linked to the dissipation and the inelastic collisions in this system.

\begin{figure*}[t!]
\centering
\begin{tikzpicture}

\node at (4.6,-.4) { \includegraphics[width=16.5cm]{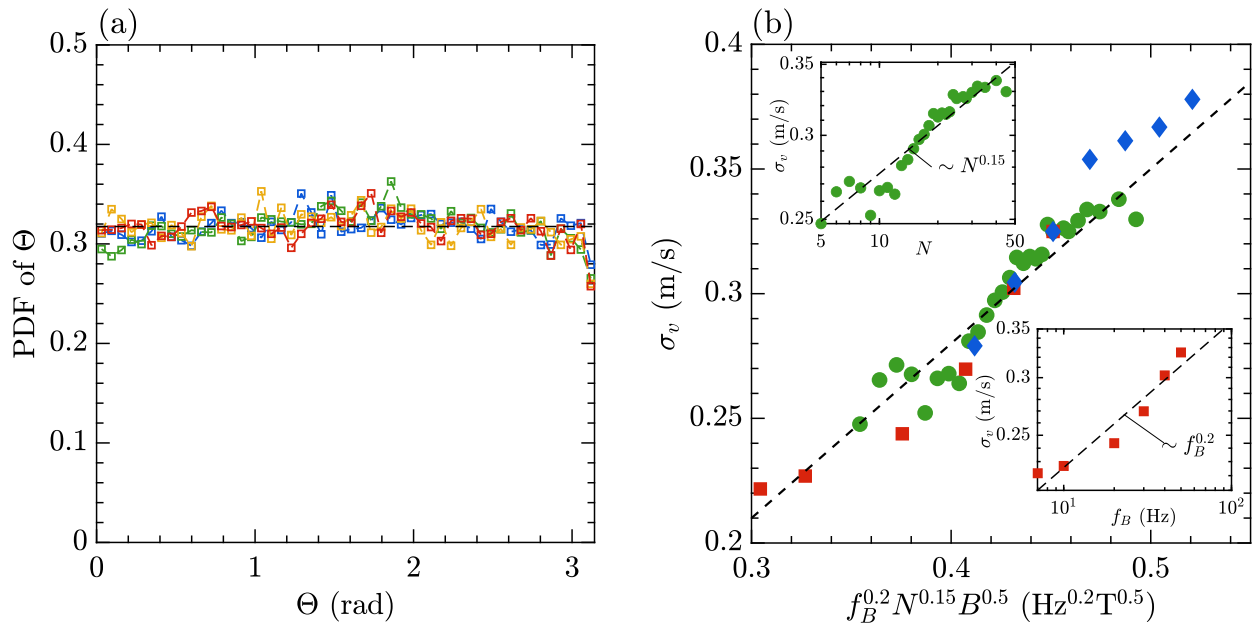}};

\end{tikzpicture}
\caption{(a) Probability density functions of the angle difference $\Theta$ for different frequencies of the magnetic field $f_B$: 7, 10, 40, and 50 Hz (cold to hot colors) \textcolor{black}{for $B=0.0162$ T and $N=25$.} (b) Main figure: Standard deviation of the linear velocity $\sigma_v$ as a function of the control parameters $(f_B, B, N)$. Top inset: $\sigma_v$ as a function of the number of magnets $N$. The dashed line represents a power law fit of $\sigma_v$ as a function of $N$. Bottom inset: $\sigma_v$ as a function of the frequency of the oscillating magnetic field $f_B$. The dashed line represents a power law fit of $\sigma_v$ as a function of $f_B$. Red squares correspond to $B=0.0162$ T and $N=25$, green circles correspond to $B=0.0162$ T and $f_B=50$ Hz, blue diamonds correspond to $f_B=50$ Hz and $N=25$.}
\label{link2}
\end{figure*}

To understand the coupling between the translational and rotational motion of the magnets, we compute the angle difference $\Theta$ between $\mathbf v$ and $\theta$. In the absence of correlations between $\mathbf v$ and $\theta$, the PDF of $\Theta$ has to be flat. Figure \ref{link2}a illustrates that the PDF of $\Theta$ is indeed flat for different frequencies of the magnetic field $f_B$. \textcolor{black}{The absence of an orientational correlation is due to the energy injection mechanism. The transfer of linear momentum during a collision between two magnets depends on the angular and linear velocity of the magnets before the collision. However, the angular momentum statistics are independent of the energy transfer during collisions, because the magnets constantly receive a magnetic torque and, therefore, quickly lose the information about the angular momentum transferred after a collision. This explains why the angular velocity is uncorrelated to the linear velocity and why the collisions between the magnets control the conversion of the rotational kinetic energy into linear kinetic energy.}  

\paragraph*{Linear kinetic energy.\textemdash} The measurements of the standard deviation of the modulus of the velocity $\sigma_v$ suggest that the relationship between the linear kinetic energy and the control parameters $\left(f_B, N, and B \right)$ is less complex than for the standard deviation of the angular velocity $\sigma_\omega$. Indeed, Fig. \ref{link2}b shows the dependence of the translational kinetic energy as a function of the control parameters $\left(f_B, N, B \right)$. The top inset in Fig. \ref{link2}b shows the dependence of $\sigma_v$ on $N^{0.15}$ and the bottom inset in Fig. \ref{link2}b shows the dependence of $\sigma_v$ on $f_B^{0.2}$. The linear energy kinetic $\sigma_v^2$ is an increasing function of $f_B, B$, similar to the rotational kinetic energy. The power law $\sigma_v \sim B^{1/2}$ is consistent with the power law measured in a 3D homogeneously forced granular gas without liquid in the container \cite{falcon2013}. The measurements also illustrate that $\sigma_v$ increases as a function of $N$. This suggests that a high number of collisions between the magnets convert more efficiently the rotational kinetic energy into translational kinetic energy.

\paragraph*{Nonequipartition of energy.\textemdash}It is also important to note that no equipartition of energy is observed in this volume-forced granular gas. The measured ratio of the rotational and linear kinetic energy is equal to $5$ because the dynamics of the magnets are driven by the input of angular momentum. This means that the magnets do not fully convert their angular momentum into linear momentum and that the kinetic energy is not equally distributed to the degree of freedom of the gas. Figure \ref{rotk2}c also suggests that the conversion of the rotational kinetic energy to linear kinetic energy is maximal at $N=16$, which corresponds to $\phi=0.07$. \textcolor{black}{We do not observe equipartition of energy here because the energy is directly injected into the magnets, which is an expected numerical result \cite{gayen2011,luding2003}.}

\paragraph*{Collision statistics.\textemdash}Since the linear kinetic energy depends on the number of magnets $N$, one can assume that the collision frequency also controls the dynamics of the magnets. We, therefore, measure the frequency of the collisions between the magnets, to compare the volume-forced granular gas with ideal gases or boundary-driven granular systems. Different dependencies on $N$ were reported. The collision frequency is proportional to $N$ in the case of ideal gases and proportional to $N^{1/2}$ in vibrated granular gases in low gravity \cite{falcon2006}.

\begin{figure*}[t!]
\centering
\begin{tikzpicture}

\node at (-4.5,-.4) { \includegraphics[width=8.5cm]{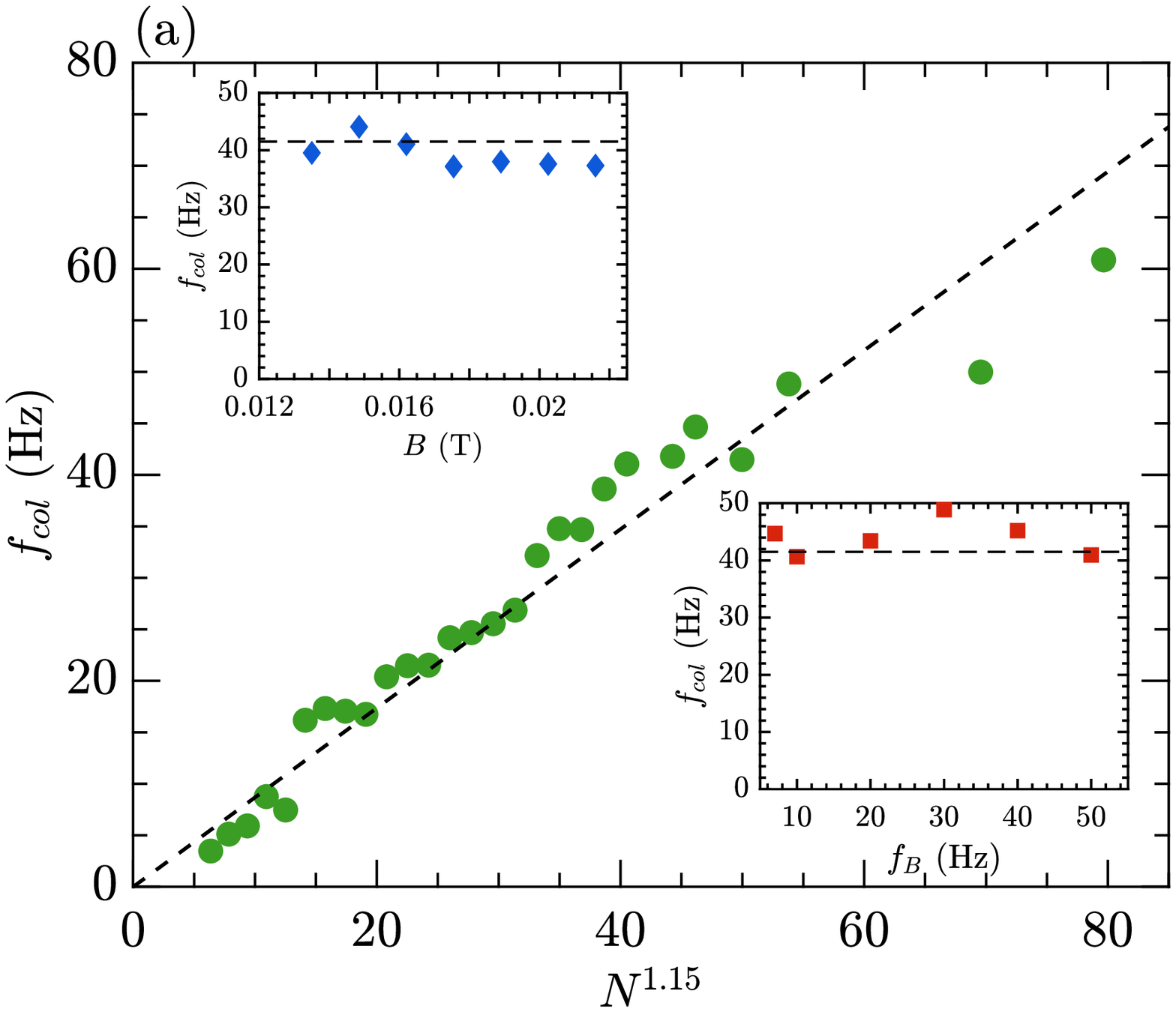}};
\node at (4.5,-.4) { \includegraphics[width=8.5cm]{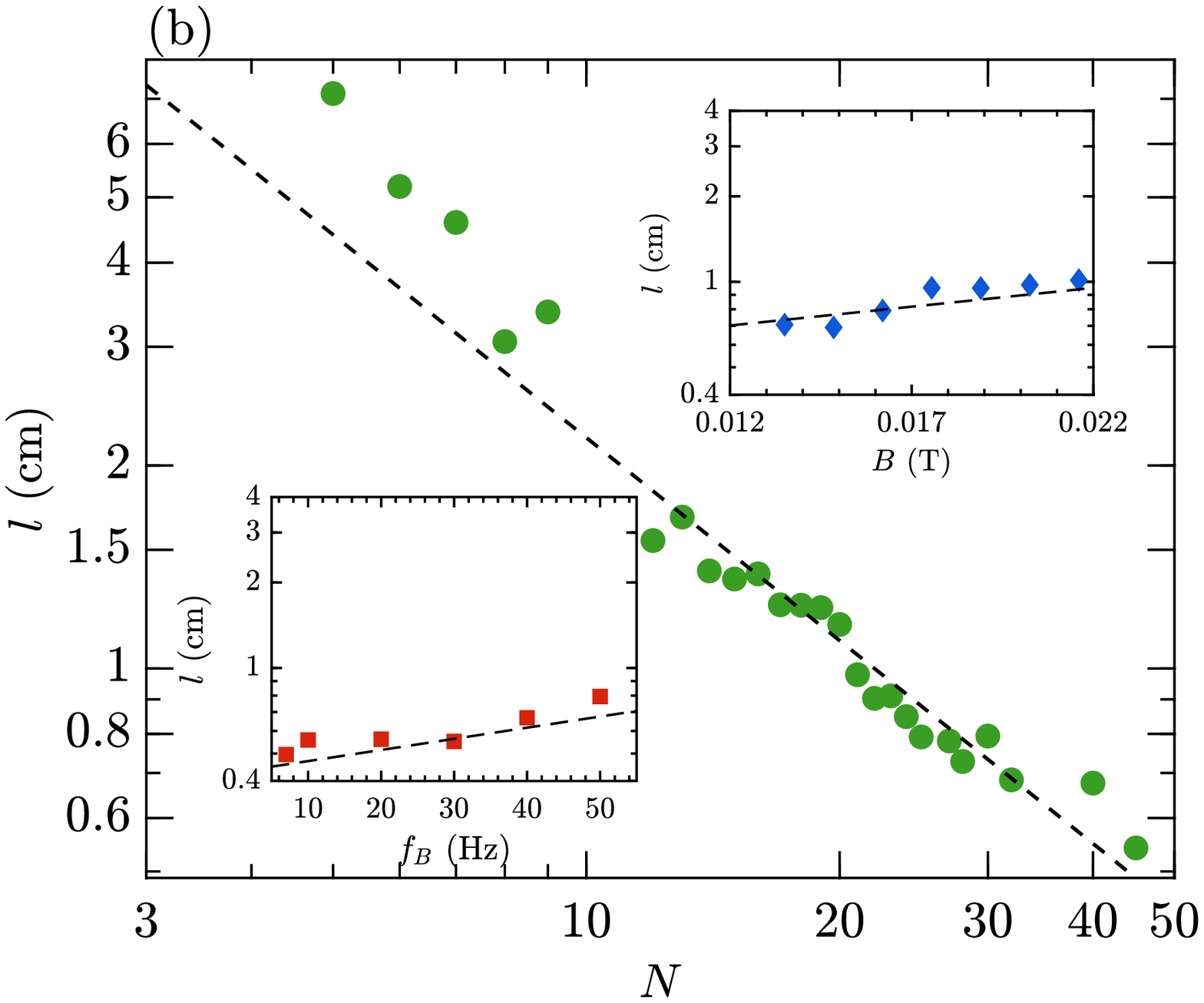}};

\end{tikzpicture}
\caption{(a) Collision frequency $f_{col}$ as a function of the number of magnets $N$. The equation of the dashed line is $y=0.87x$. Top inset: Collision frequency $f_{col}$ as a function of the intensity of the magnetic field $B$. Bottom inset: Collision frequency $f_{col}$ as a function of the frequency of the oscillating magnetic field $f_B$. (b) Mean free path $l$ as a function of the number of magnets $N$. The equation of the dashed line is $y=22/x$. Top inset: mean free path $l$ as a function of the intensity of the magnetic field $B$. Bottom inset: mean free path $l$ as a function of the frequency of the oscillating magnetic field $f_B$. In both figures, red squares correspond to $B=0.0162$ T and $N=25$, green circles correspond to $B=0.0162$ T and $f_B=50$ Hz, blue diamonds correspond to $f_B=50$ Hz and $N=25$.}
\label{meanfree}
\end{figure*}

Figure \ref{meanfree}a shows that the collision frequency $f_{col}$ solely depends on the number of magnets $N$. The relationship between the collision frequency and the number of magnets is $f_{col}\sim N^{1.15}$, as illustrated by the dashed line in Fig. \ref{meanfree}a. This result is slightly different from the linear relationship between $f_{col}$ and $N$ in the case of ideal gases. The top and bottom insets in Fig. \ref{meanfree}a show that the collision frequency $f_{col}$ is independent of both $f_B$ and $B$. Such measurements imply that any change in the kinetic energy does not affect the frequency of the collision. This quantity is, therefore, solely controlled by the number of magnets.

\textcolor{black}{The average distance a particle travels between collisions with other moving particles, the mean free path $l$, can be estimated from the following expression}

\begin{equation}
l=\sigma_v/f_{col}
\end{equation}

\begin{table*} 
\begin{ruledtabular}
\begin{tabular}{ccccc}
& Volume forced& Volume forced &Ideal gas&Boundary forced\\
&3D gas (dissipative) & 2D gas (dissipative) & (non dissipative) &granular gases (dissipative)\\
&\cite{falcon2013} & Present study & \cite{reif2009} & \cite{falcon1999b,falcon2006} \\
\hline
& & & \\
Injection of energy & Volume & Volume & - & Surface ($\mathcal{V}$\textsubscript{wall}) \\
Kinetic energy $\sigma_v^2$ & $E_k \sim B$ & $E_k \sim f_B^{0.4} N^{0.3} B$ & $E_k \sim T$ & $E_k \sim \mathcal{V}_{\text{wall}}^{\alpha\left(N\right)}$ \\
& & & \\
Equation of state & $PV \sim N E_k V_p / V$ & - & $PV \sim N E_k$ & $PV \sim E_k$ \\
& & & \\
Equipartition of energy & No & No & Yes & Sometimes \\
& & & \\
Clustering & No & No & No & Yes \\
& & & \\
Linear velocity& Exponential & Stretched exponential & Maxwell-Boltzmann & Stretched exponential \\
distributions & Independent of $N$ & with $\alpha=1.6$& Independent of $N$ & with $\alpha\left(N\right) \in \left[0,2\right]$ \\
& & Independent of $N$ & & Depends on $N$\\
Collision frequency $f_{col}$& $N B^{1/2}$ & $N^{1.15}$ & $N T^{1/2}$ & $\sqrt{N} \mathcal{V}_{\text{wall}}$ \\
Mean free path $l$ & - & $1/N$ & $1/N$ & $1/\sqrt{N}$ \\
\end{tabular}
\end{ruledtabular}
\caption{Comparison between volume-forced gases of magnets (3D or 2D), the ideal gas, and boundary-forced granular gases.Volume-forced 3D gas: Measurements performed using an accelerometer clamped on the lid of a 3D granular gas \cite{falcon2013}. Volume-forced 2D gas (present study):  Measurements using Lagrangian tracking techniques in a quasi-2D water-filled cell. For boundary-forced granular gases (fourth column) $\alpha\left(N\right) \in \left[0,2\right]$ is a coefficient that depends on the number of particles $N$ in the granular gas. \textcolor{black}{$\mathcal{V}_{\text{wall}}=2 \pi F A$ is the velocity of the walls of the container, with $A$ the amplitude of the oscillations and $F$ their frequency.}}
\label{table1}
\end{table*}

One would expect the mean free path to be proportionate to $l=f_B^{0.2}B^{0.5}/N$ because $\sigma_v$ is proportionate to $f_B^{0.2}N^{0.15}B^{0.5}$ (Fig. \ref{link2}b), and $f_ {col}\sim N^{1.15}$ (Fig. \ref{meanfree}a). Figure \ref{meanfree}b shows that the mean free path $l$ is indeed inversely proportionate to the number of magnets $N$, as illustrated by the dashed line. The top and bottom insets in Fig. \ref{meanfree}b show the weak dependences of the mean free path $l$ on $f_B$ and $B$. The Knudsen number can be defined as $K = l/L$, with $L=\sqrt{dh}\approx11$ cm with $d=15$ cm the length of the container and $h=8$ cm the height of the container. The measurements show that the Knudsen number is in the range of $ \left[0.05,0.64\right]$ and is inversely proportional to $N$ for our range of parameters, implying that the gas is in a kinetic regime. \textcolor{black}{However, a slight departure from the dashed line is observed for $N<10$ ($\phi<0.045)$ in Fig. \ref{link2}}. The value of the Knudsen number illustrates that the dynamics of the gas are solely dominated by the collisions between the magnets and not by the collisions between the magnets and the boundaries. This confirms that the collisions between the magnets transfer linear kinetic energy to the gas and control the dynamics. \textcolor{black}{In addition, Fig. \ref{link}c illustrates that the shape of the distribution of the horizontal velocity is independent of $N$ and also suggests no transition towards a Knudsen regime.}

\paragraph*{Discussion.\textemdash} The results of the measurements performed using Lagrangian tracking techniques in this volume-forced granular gas are fundamentally different from the ones performed using other 3D dry granular gas systems, as summarized in table \ref{table1}. The results strongly emphasize the difference in the dependence on the number of particles $N$ between different experimental systems, such as boundary-driven granular gases. Here, even though the shape of the distribution of the angular velocity changes as a function of the magnetic field frequency, the shape of the distribution of the linear velocity does not change. The kinetic energy in the volume-forced granular gas depends on $N$, and the linear velocity distributions also resemble the heavy-tailed distributions seen in boundary-forced granular gases. The collision frequency depends on $N$ with a larger exponent in the volume forced granular gas than for ideal gases or boundary-driven gases, probably because the kinetic energy depends on $N$. The mean free path is measured using the relationship between the linear kinetic energy and the collision frequency and implies that the mean free path depends on $N$ but also on $f_B$ and $B$. However, we do not have an equation of state in this system because we did not measure the pressure independently from the trajectories.

\paragraph*{Conclusion.\textemdash}We present the first statistical measurements of a 2D homogeneously-forced granular gas driven stochastically by injecting rotational kinetic energy into magnets, extending the measurements performed in 3D \cite{falcon2013}. This system differs from previous experimental studies of granular gas where the energy is injected by vibrating the boundaries. We report the differences between this homogeneously-forced granular gas, ideal gas, and non-equilibrium boundary-forced dissipative granular gas. Here, we do not observe any cluster formation, even for a large volume fraction $\phi=0.21$, nor the equipartition of the energy. The velocity distributions are stretched exponentials, as for other 3D boundary-forced dry granular gas systems, but the exponent does not depend on $N$ \textcolor{black}{and} is close to the value of 3/2 derived theoretically \cite{van1998}.

Even though the shape of the distributions of the angular velocities changes, the gas dynamics are solely controlled by the average linear kinetic energy. The homogeneously-forced granular gas studied here presents interesting physical properties which can be useful in medical applications if the magnets are scaled down to a nanometric scale \cite{fortin2007,gazeau2008}. The generated flow also offers opportunities for improved chemical mixing or studying the pair dispersion, and diffusion of passive scalars in turbulent flows \cite{cazaubiel2021,gorce2022}.

\begin{acknowledgments}

We thank A. Di Palma and Y. Le Goas for technical help. This work has been supported by ESA Topical Team on granular materials No 4000103461 and from the Simons Foundation MPS No 651463-Wave Turbulence grant (USA).

\end{acknowledgments}

\appendix*

\section{Angular velocity distributions for different $N$ }

\begin{figure}[h!]
\centering
\begin{tikzpicture}

\node at (-4.5,-.4) { \includegraphics[width=8.5cm]{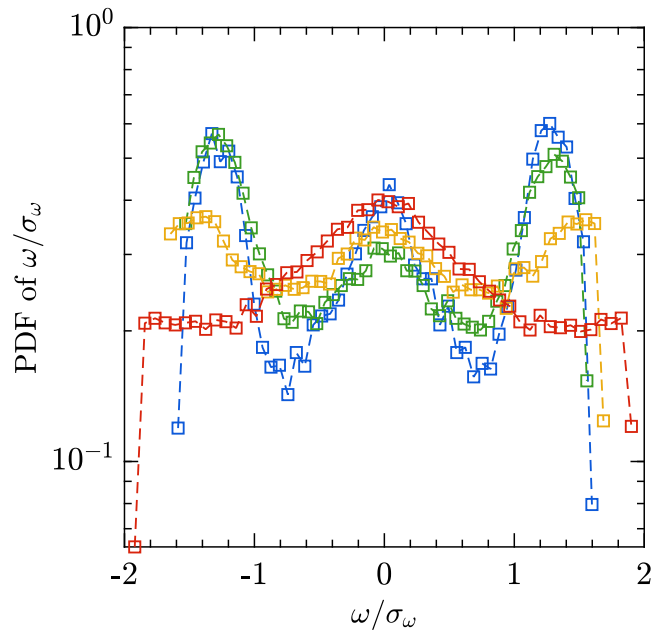}};

\end{tikzpicture}
\caption{
\textcolor{black}{
 Probability density functions (PDFs) of the rescaled angular velocity $\omega/\sigma_\omega$ for different number of magnets $N$: 11, 20, 30, and 45 (cold to hot colors). The frequency of the magnetic field is equal to $f=50$ Hz and the intensity of the magnetic field to $B=0.0162$ T. }
}
\label{app}
\end{figure}

\textcolor{black}{
Measurements of the distributions of the angular velocity of the magnets as a function of the number of magnets $N$ are shown in Fig. \ref{app}. The double humps are observed for $N=30$ ($\phi=0.135$) but not for $N=45$  ($\phi=0.21$). This can be explained because of the high density of magnets for $N=45$, and, therefore, the high number of collisions (Fig. \ref{meanfree}a), prevents the magnets from synchronizing with the vertical oscillating magnetic field.
}

\end{document}